\documentclass[conf]{new-aiaa}
\usepackage[utf8]{inputenc}
\usepackage{graphicx}
\usepackage{amsmath}
\usepackage{bm}
\usepackage{epstopdf}
\usepackage{cleveref}
\usepackage{multirow}
\usepackage{multicol}
\usepackage{subfig}     
\usepackage{units}
\usepackage[version=4]{mhchem}
\usepackage{siunitx}
\usepackage{longtable,tabularx}
\usepackage{xcolor}

\title{Resolvent Analysis of an Under-expanded Planar Supersonic Impinging Jet}
\author{Qiong Liu\footnote{Present address: Department of Aerospace Engineering, UIUC. Email:qiongl@illinois.edu}, Chitrarth Prasad\footnote{Postdoctoral Researcher, Department of Mechanical and Aerospace Engineering} and Datta V. Gaitonde\footnote{John Glenn Chair Professor, Department of Mechanical and Aerospace Engineering}}
\affil{The Ohio State University, OH 43210, USA}

\begin{document}

\maketitle

\begin{abstract}
{\color{blue}This investigation aims to assess the effect of different types of actuator forcing on the feedback loop of an under-expanded Mach 1.27 planar impinging jet using a resolvent framework.
To this end, we employ a Large Eddy Simulation database as a truth model.
The time and spanwise-averaged mean flow is taken as an input to global stability and resolvent analyses with the purpose of examining both the intrinsic instability and input-output characteristics. 
The results show that the inherent instability and primary energy amplification are attributed to the Kelvin-Helmholtz (K-H) instability. 
Moreover, the K-H response modes obtained from the resolvent analysis are in reasonable agreement with Spectral Proper Orthogonal Decomposition (SPOD) modes from the unsteady LES data.
Insights into noise control are obtained by localizing the actuator forcing to the nozzle lip and the ground plate by imposing component-wise forcing to mimic different notional actuators.
It is observed that energy amplification obtained for the localized component-wise forcing is different from the global resolvent analysis and dependent on the type of actuator. 
This provides insights into the type, wavenumber and frequency of actuators for active flow control.}

\end{abstract}

\section*{Nomenclature}

{\renewcommand\arraystretch{1.0}
\noindent\begin{longtable*}{@{}l @{\quad=\quad} l@{}}
$\rho$ & Density\\
$u$ & Streamwise velocity\\
$T$ & Temperature\\
$a_\infty$ & speed of sound\\
$c$ & Local speed of sound \\
$D$ & Distance between nozzle walls\\
$\bm{B}$ & Hydrodynamic FT Mode \\
$\psi'$ & Irrotational Scalar Potential \\
$\psi_a'$ & Acoustic Scalar Potential \\
$\psi_t'$ & Thermal Scalar Potential \\
$S$ & Entropy \\
$St$ & Oscillation frequency \\
$\beta$ & Spanwise wavenumber \\
$\mathbb{R}$ & Resolvent operator\\
$\mathbb{Q}=[\hat{\bm q}_1, \hat{\bm q}_2,\dots, \hat{\bm q}_n]$ & response modes\\
$\mathbb{F}=[\hat{\bm f}_{1}, \hat{\bm f}_{2},\dots, \hat{\bm f}_{n}]$ & forcing modes \\
\multicolumn{2}{@{}l}{Subscripts} \\
$j$ & Jet exit quantities \\
$\infty$ & Ambient quantities
\end{longtable*}}

\section{Introduction}
The study of a supersonic jet impinging on a flat plate is of great practical relevance for the design of V/STOL aircraft and launch vehicles. 
The interaction of the jet with the impingement surface produces a very complex flow-field, which can be divided into a free jet region, an impingement zone and a wall jet region.
In addition to the acoustic radiation from the free-jet and its subsequent reflection from the impingement surface, the extra noise sources due to the complex flow-field at the impinging zone and the ensuing wall jet often result in the flow-field being dominated by a self-reinforcing aeroacoustic feedback loop~\cite{powell1988sound}.
This feedback process has been the subject of several investigations, a comprehensive review of which can be found in Ref.~\cite{edgington2019aeroacoustic}.
The impingement of the jet on the plate produces upstream travelling acoustic waves that propagate to the nozzle exit and provide a periodic forcing to the convectively unstable thin shear layer.
These perturbations grow into large scale structures that convect downstream and impinge on the surface resulting in acoustic waves at a suitable phase and frequency which complete the feedback loop~\cite{prasad2021exchange}.
The acoustic tones produced by this feedback phenomenon result in significant fluctuating loads that can be up to 10dB louder than their free-jet counterparts~\cite{krothapalli1999flow}. 

Both active and passive control techniques have been investigated to target the individual elements of this feedback loop. 
Passive control techniques focus on modifying the impingement geometry itself so as to avoid these acoustic waves from perturbing the shear layer.
These designs include, but are not limited to, the use of oblique impingement \cite{worden2013acoustic,brehm2016noise}, the installation of front-covers within the vertical flame path \cite{tsutsumi2015study} and the introduction of a wall with a circular cut-out to block the upstream traveling waves from the impingement area towards the nozzle lip~\cite{prasad2021effectlaunchpad}.
Active noise reduction methods on the other hand, often focus on fluid injection at certain strategic locations around the nozzle exit and beneath the impingement plate~\cite{alvi2003control,norum2004reductions,fukuda2011examination,salehian2018numerical}.
Although these approaches have enjoyed some measure of success, a more systematic study of the sensitivity of an impinging jet flow-field to different types of actuators forcing is lacking.

Active control techniques are more attractive than their passive counterparts, as they can be operated on-demand. 
The most general parameters in the design of any active noise control methodology is the type and spatial distribution (wavenumber) of the actuators, and the amplitude and frequency of actuation. 
The goal of the present study is assess the effect of choosing different types of actuator forcing and location on the feedback loop of an under-expanded Mach 1.27 planar impinging jet using a resolvent framework.
This is motivated by the promising success of resolvent analysis in controlling the feedback process in cavity flows~\cite{liu2018resolvent,liu2021unsteady}.
Although the mean flow needed for resolvent analysis could be obtained from a Reynolds Averaged Navier-Stokes (RANS) solution; we employ a carefully computed Mach~1.27 LES database to serve as a truth model and therefore use the same database to obtain the mean flow for analysis purposes. 
A detailed description of the LES database is provided in Section~\ref{Sec:LES}. Section~\ref{sec:MA} provides a brief discussion of the resolvent framework used in this study. 
Some preliminary results are presented in Section~\ref{sec:Results}.
A summary of ongoing work and concluding remarks are given in Section~\ref{conclusion}.

\section{LES Database}\label{Sec:LES}
The LES database used in the present study consists of an unheated planar under-expanded Mach 1.27 jet impinging on a flat plate placed at 4 jet diameters from the nozzle exit.
This impingement height is chosen based on observed resonant mode sensitivity to nearby heights in previous single impinging jet studies~\cite{Bhargav2021,Davis2015,Kumar2013}.
The database is generated by solving the full 3D compressible Navier-Stokes equations in generalized curvilinear coordinates using a well-validated inhouse solver SAFF. 
A third-order upwind biased Roe scheme~\cite{roe1981approximate} in conjunction with the van Leer harmonic limiter~\cite{van1979towards} is employed for the spatial discretization of the inviscid fluxes. 
The viscous fluxes are computed using a second-order central differencing scheme, whereas the time integration is achieved by using an implicit second-order diagonalized~\cite{pulliam1981diagonal} Beam Warming scheme~\cite{beam1978implicit}.

\begin{figure}
    \centering
    \includegraphics[width=0.9\textwidth]{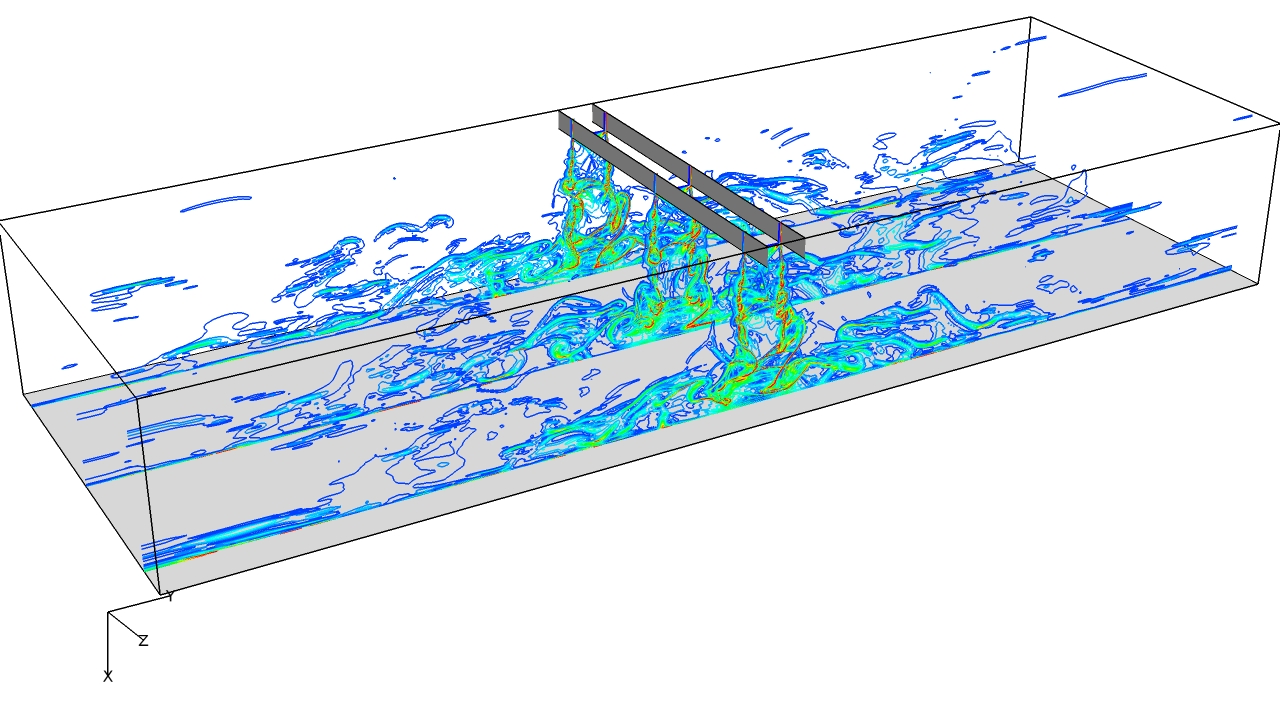}
    \caption{A schematic of the computational domain for LES calculations. The geometry consists of two parallel walls at distance of 1 inch with a sonic inflow.}
    \label{fig:domain}
\end{figure}

A schematic of the computational domain is presented in Fig.~\ref{fig:domain}.
The geometry consists of two parallel walls separated at a distance $D$ of 1 inch covering the entire spanwise extent of the domain.
A cartesian coordinate system is chosen. with the jet exhausting in the $x-$direction.
The nozzle exit is located at $x=0.5D$, whereas the plate is located at $x=4.5D$.
The domain extends up to $15D$ in the $y-$direction on either side of the jet axis, whereas the spanwise extent of the domain ($z-$direction) is $10D$.
The calculations are performed in a structured mesh containing 521, 917 and 81 points in the $x$, $y$ and $z$ directions respectively.
Non-reflecting freestream boundary conditions are applied at all the outer boundaries.
Iso-thermal wall conditions with $T_w=1.12T_\infty$ are applied at all solid boundaries including the nozzle walls and the impingement plate.
The inflow consists of a laminar boundary layer profile with thickness $\delta= 0.1D$ following the approach described by Bogey and Bailly~\cite{bogey_bailly_2010} to match the onset of shear layer turbulence observed in experiments~\cite{nataraj2020unsteady}.
This manner of inflow specification has proven effective in replicating the experimentally observed shear layer development in both single and twin impinging jets~\cite{stahl2021distinctions}.

All flow quantities are used in their dimensionless form. 
The jet exit conditions, $U_j = 343.7 \unit{m/s}, T_j = 293 \unit{K} \text{ and } \rho_j=1.2061 \unit{kg/m^3}$ are used as the reference quantities for velocity, temperature and density respectively. 
The nozzle exit diameter, $D=0.0254m$ is used as the reference length, whereas the pressure is normalized by $\rho_j {U_j}^2$. 
The ambient temperature is $T_\infty = 293 \unit{K}$. 

\begin{figure}[!h]
    \centering
    \includegraphics[width=0.8\textwidth]{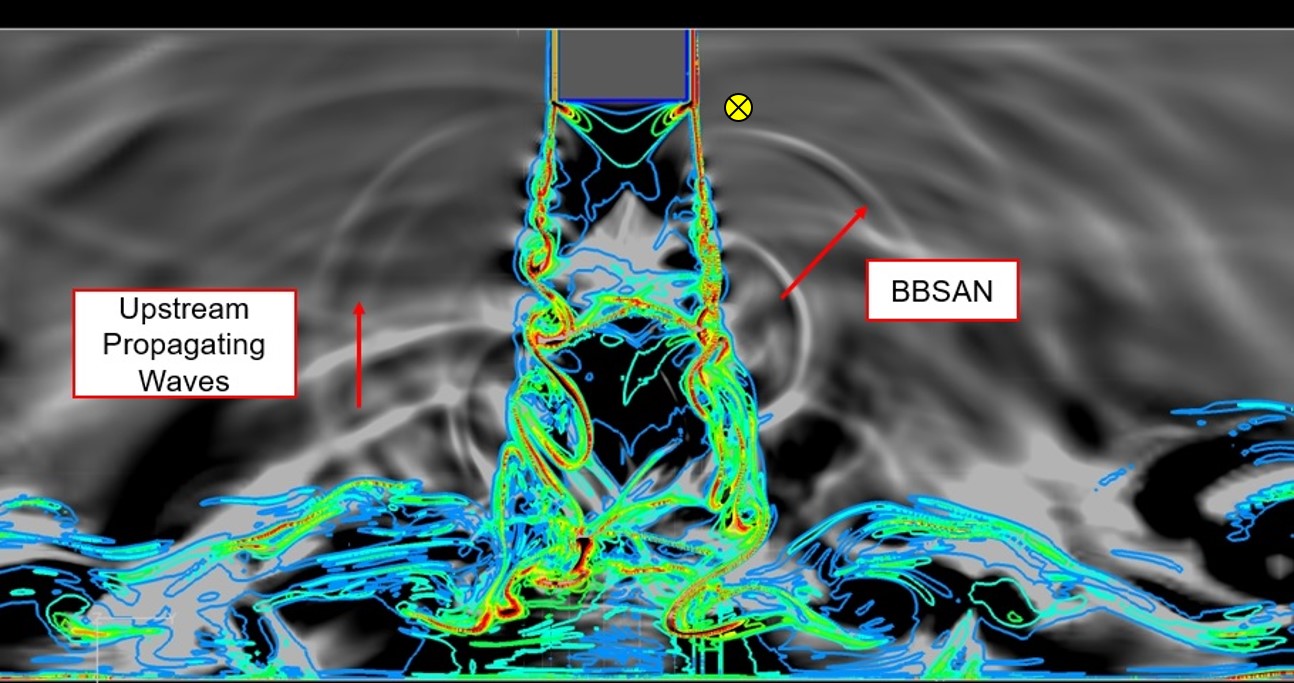}
    \caption{Density gradient magnitude contours with background dilatation contours in a 2D streamwise plane.}
    \label{fig:dengrad}
\end{figure}

Figure~\ref{fig:dengrad} shows the density gradient magnitude contours (color) superimposed on velocity dilatation contours (gray-scale) at an arbitrary time step in a 2D streamwise plane. 
The density gradient contours illustrate the turbulent structures and the shock cells in the jet plume, whereas the background dilatation contours give a qualitative picture of the acoustic waves generated due to this mixing. 
Two distinct types of sound waves can be identified: the broadband shock associated noise in the sideline and upstream direction at locations where the shock cells interact with the shear layer and the upstream travelling waves due to the impinging zone and the ensuing supersonic wall jet.

\begin{figure}
    \centering
    \includegraphics[width=0.75\textwidth]{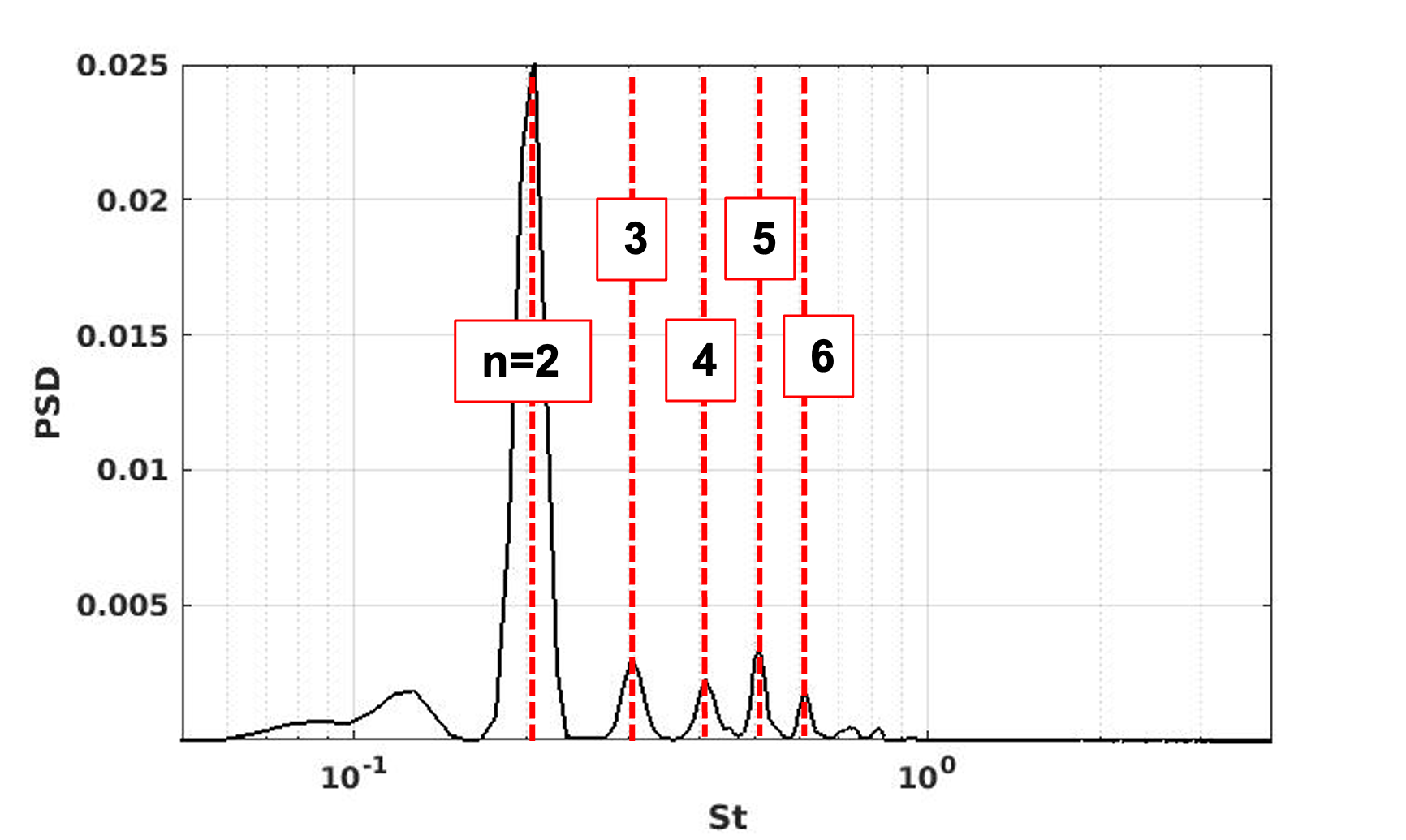}
    \caption{Power Spectral Density of near-field pressure. The impinging tone harmonics $n=2$ to $6$ are identified.}
    \label{fig:PSD}
\end{figure}

Figure~\ref{fig:PSD} shows the power spectral density of pressure as a function of Strouhal number ($St=fU_j/D$) at the point marked by yellow circle in Fig.~\ref{fig:dengrad}. 
The high-amplitude discrete frequency feedback tones due to the feedback loop between the nozzle exit and the ground plate are clearly visible in the spectrum.
In order to validate these impinging tone frequencies, we make use of Powell's feedback model.
Following Powell~\cite{Powell1953}, the fundamental impingement tone frequency $F$, and its harmonics $n$ are commonly expressed as
\begin{equation}
\label{afbl}
\frac{n}{F}=\frac{H}{U} + \frac{H}{a} +p,
\end{equation}
where $H$ represents the height of impingement, $U$ is the average shear layer convection speed, and $a$ is the speed of sound.
The time lag in the receptivity and sound generation components is accounted for in the phase lag term \textit{p} in order to recover the observed acoustic tones.
In the present investigation, the speed of the downstream convecting coherent structures and upstream propagating acoustic waves are simultaneously measured by a line of simulated probes along the jet shear layer from the nozzle lip to the impingement plate (not shown).
The pressure fluctuations at every point along the shear-layer line of probes is cross-correlated with that at the nozzle exit location to calculate the shear layer convection speed. 
The speed of sound is assumed to be $343.7$ m/s, resulting in an equivalent fundamental frequency equal to $St=0.10$. 
This approach sets the phase lag term $p$ in equation~\ref{afbl} to zero by assuming it is accounted for in the computed average shear-layer convection speed.
Using this key result the impinging tones observed in Fig.~\ref{fig:PSD} can be identified as harmonics $n=2$ to $6$ of the fundamental frequency.
This provides confidence in the use of this database in the present work.
A detailed calculation of the individual elements of the feedback loop will be presented in the final manuscript.

The LES flow-field is averaged in both time and spanwise direction over $300D/U_j$ time snapshots to provide an input for the global stability and resolvent analyses. 
A brief summary of these techniques is presented next.

\section{Modal analysis}\label{sec:MA}
\subsection{Global stability analysis}\label{sec:Stability}
An adoption of Reynolds averaging in the turbulent flow separates the mean Reynolds stresses and fluctuating Reynolds stress. 
The mean Reynolds stresses are implicitly included in the turbulent mean flow that are obtained from the LES. 
In comparison, the fluctuating Reynolds stresses are amplified/diminished at certain frequencies, which is assumed as a nonlinear forcing acting on the large-scale structures in the turbulent flow. 
This point of view is similar to the one used in Refs~\cite{farrell1993stochastic,mckeon2010critical} for modal analyses of high Reynolds number turbulent flows and has been shown to provide significant insights in understanding flow dynamics and informing flow control. 
In the present study, we adopt this methodology by first neglecting the nonlinear forcing and assessing the unforced behavior of the linear modes based on a time averaged impinging jet flow field. This information is then utilized to assess the response of the flow to non-linear forcing as shown later.

The instantaneous flow variables are Reynolds decomposed into a time-averaged flow state $\bar{\bm{q}}\equiv[\bar{\rho},\bar{u},\bar{v},\bar{w},\bar{T}]$ and a statistically stationary fluctuating components $\bm{q}'\equiv[\rho',u',v',w',T']$. By substituting the Reynolds decomposed state variables into the Navier–Stokes (NS) equations, the governing equation for the fluctuation $\bm{q}'$ becomes
\begin{equation}
\label{eqn:decom_NS}
    \frac{\partial \bm q'}{\partial t}=\mathbb{L}(\bar{\bm q})\bm q',
\end{equation}
where $\mathbb{L}(\bar{\bm q})$ is the Navier-Stokes operator linearized about the time-averaged flow state $\bar{\bm{q}}$. 

Since the present planar impinging flow is periodic in the spanwise direction, 
the fluctuation term $\bm q'$ can be expressed as Fourier modes with real a spanwise wavenumber $\beta$ and a complex frequency $\lambda$ as follows
\begin{equation}
\label{eqn:fouriermodes}
\bm q' (x,y,z,t)={\hat{\bm q}}(x,y)e^{\text{-i}(\beta z + \lambda t)}.
\end{equation}
Here the spanwise wavenumber is normalized by $D$ and
the complex value of $\lambda$ is non-dimensionalized using $D/2\pi U_j$. 
The real part $\lambda_r$ represents the oscillation frequencies and imaginary part $\lambda$ denotes the growth ($\lambda>0$) or damping rates ($\lambda<0$) of linear modes.  
Substituting the modal expression of $\bm q' (x,y,z,t)$ from eqn.~\ref{eqn:fouriermodes} into eqn.~\ref{eqn:decom_NS} yields an eigenvalue problem of the linear system
\begin{equation}
\label{eqn:eigenvalue}
\mathbb{L}(\bar{\bm q}, \beta)\hat{\bm q}=\lambda \hat{\bm q}.
\end{equation}

Given the time and spanwise averaged flow with prescribed spanwise wavenumbers, this large eigenvalue problem can be solved to obtain linear modes in terms of their eigenvalues and eigenvectors $(\lambda, \hat{\bm q})$. $\mathbb{L}$ represents the system of the linearized compressible continuity, Navier-Stokes, and energy equations for an ideal gas in Cartesian coordinates. 

For computational efficiency, we construct the linear operator $\mathbb{L}$ using a smaller computational domain and reducing the grid relative to the one used in LES. 
This computational domain extents from $-9D \leq y \leq 9D$ and is sufficient to illustrate the primary flow physics of interest that occur between the jet nozzle and impingement surface.
At inflow, Dirichlet boundary condition is used on all perturbation variables. 
At the outflow, sponge zones with width of $3D$ are prescribed. The spatial discretion uses spectral element methods as described in Refs.~\cite{liu2016linear,liu2021acoustic} with element size of $1327$ and polynomial order of $5$. 
The eigenvalue problem is solved using an efficient shift-and-invert Arnoldi algorithm with six solved eigensolution at Krylov space dimension of 16 and a residual tolerance of $10^{-7}$ at each shift value. 
The results converge to at least seven significant figures and are verified to be independent of the domain size and mesh resolution.

\subsection{Resolvent analysis}\label{sec:Resolvent}
The input-output characteristics of the impinging jet flow are obtained from a resolvent analysis. 
We now retain the effect of the nonlinear forcing and consider its effect on the flow through the resolvent operator. 
The resulting fluctuation NS equations can be expressed as an input-output system~\citep{jovanovic2005componentwise,mckeon2010critical,schmid2012stability} given by,
\begin{equation}
\label{eqn:decom_NS1}
    \frac{\partial \bm q'}{\partial t}=\mathbb{L}(\bar{\bm q})\bm q'+\mathbb{M}\bm f'
\end{equation}
where $\mathbb{L}(\bar{q})$ is the Navier-Stokes operator linearized about the base state $\bar{\bm{q}}$,
the finite-amplitude nonlinear terms are incorporated in ${\bm f}'$, and 
$\mathbb{M}$ is a coupling matrix and is discussed later.

We use of a similar modal ansatz with spanwise wavenumber $\beta$ and temporal frequency $\lambda$ for both the state vector fluctuations and the forcing as shown below
\begin{equation}\label{eqn: ansatz}
\begin{array}{c}
\bm{q}'(x,y,z,t)=\hat{\bm{q}}(x,y)e^{-\text{i}(\lambda t+\beta z)}, \\ 
\bm{f}'(x,y,z,t)=\hat{\bm{f}}(x,y)e^{-\text{i}(\lambda t+\beta z)}.
\end{array}
\end{equation}
Inserting these into eqn.~\ref{eqn:decom_NS1}, we obtain     
\begin{equation}\label{eqn: Resolvent}
	\hat{{\bm q}} 
	=[-\text{i}\lambda \mathbb{I} - \mathbb{L}({\bm{\bar q}},\beta)]^{-1}\mathbb{M}\hat{{\bm f}},
\end{equation}
where $\mathbb{R}=[\text{i}\lambda \mathbb{I} - \mathbb{L}({\bm{\bar q}},\beta)]^{-1}\mathbb{M}$ is the resolvent operator, which serves as a transfer function between the input $\hat{\bm f}$ and the corresponding output $\hat{\bm q}$ for a given flow state ($\bm{\bar{q}}$) and modal parameters ($\beta$ and $\lambda$). 

The energy amplification of the system is evaluated from the ratio of output to input energy $\frac{||\hat{\bm{q}}||_E}{||\hat{\bm{f}}||_E}$, where $||\cdot||_E$ is an energy norm. 
A proper selection of the energy norm is crucial in order to prevent divergence in non-modal results. 
Here we use the compressible energy \emph{Chu}-norm~\citep{chu1965energy} given by
\begin{displaymath}
E=\int_S \left[ 
	\frac{\bar{a}^2 \rho^2}{\gamma \bar{\rho}}+\bar{\rho}({u}^2+{v}^2+{w}^2) + \frac{\bar{\rho}C_v T^2}{\bar{T}}
	\right] {\text d}s
\end{displaymath}
where $S$ is the domain of interest. 
A singular value decomposition (SVD) of the resolvent operator facilitates a ranking of the energy amplification ratio in descending order. 
This yields
\begin{equation}\label{eqn: Resolvent_2}
	W^{\frac{1}{2}}\mathbb{R}W^{-\frac{1}{2}} = \mathbb{Q} \Sigma  \mathbb{F}^\ast,
\end{equation}
in terms of the weight matrix, $W$, based on the compressible energy norm $E$. 
The matrix $\mathbb{Q}=[\hat{\bm q}_1, \hat{\bm q}_2,\dots, \hat{\bm q}_n]$ holds the set of optimal response directions and $\mathbb{F}=[\hat{\bm f}_{1}, \hat{\bm f}_{2},\dots, \hat{\bm f}_{n}]$ contains the corresponding forcing directions, where $\hat{\bm q}_i=(\hat{\rho}_r,\hat{u}_r,\hat{v}_r,\hat{w}_r,\hat{T}_r,)$ and $\hat{\bm f}_i=(\hat{\rho}_f,\hat{u}_f,\hat{v}_f,\hat{w}_f,\hat{T}_f,)$ with $n$ is number of solved singular values.
The superscript~$\ast$ denotes the Hermitian transpose. 
The singular values $\Sigma=\text{diag}(\sigma_1,\sigma_2,\dots,\sigma_n)$ represent the energy amplification (gain) between response and forcing modes.

\section{Results and discussions} \label{sec:Results}
\subsection{Global linear modes}
\begin{figure}
\centering
\includegraphics[width=0.9\textwidth]{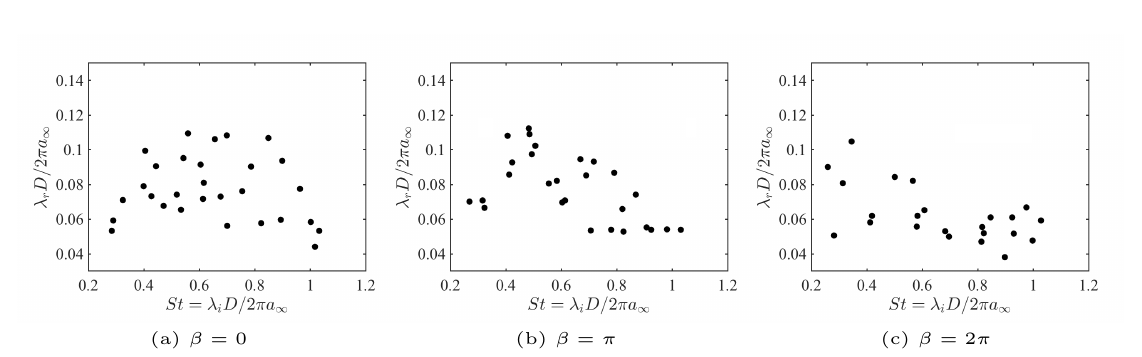}
\caption{Eigen-spectra of the time and spanwise-mean base flow at $\beta=0$, $\pi$ and 2$\pi$. The blue and black circles are used to indicate the eigenvalues of global mode shown in figure ~\ref{fig:globalmodes}. }
\label{fig:figure1}
\end{figure}

\begin{figure}
\centering
\includegraphics[width=\textwidth]{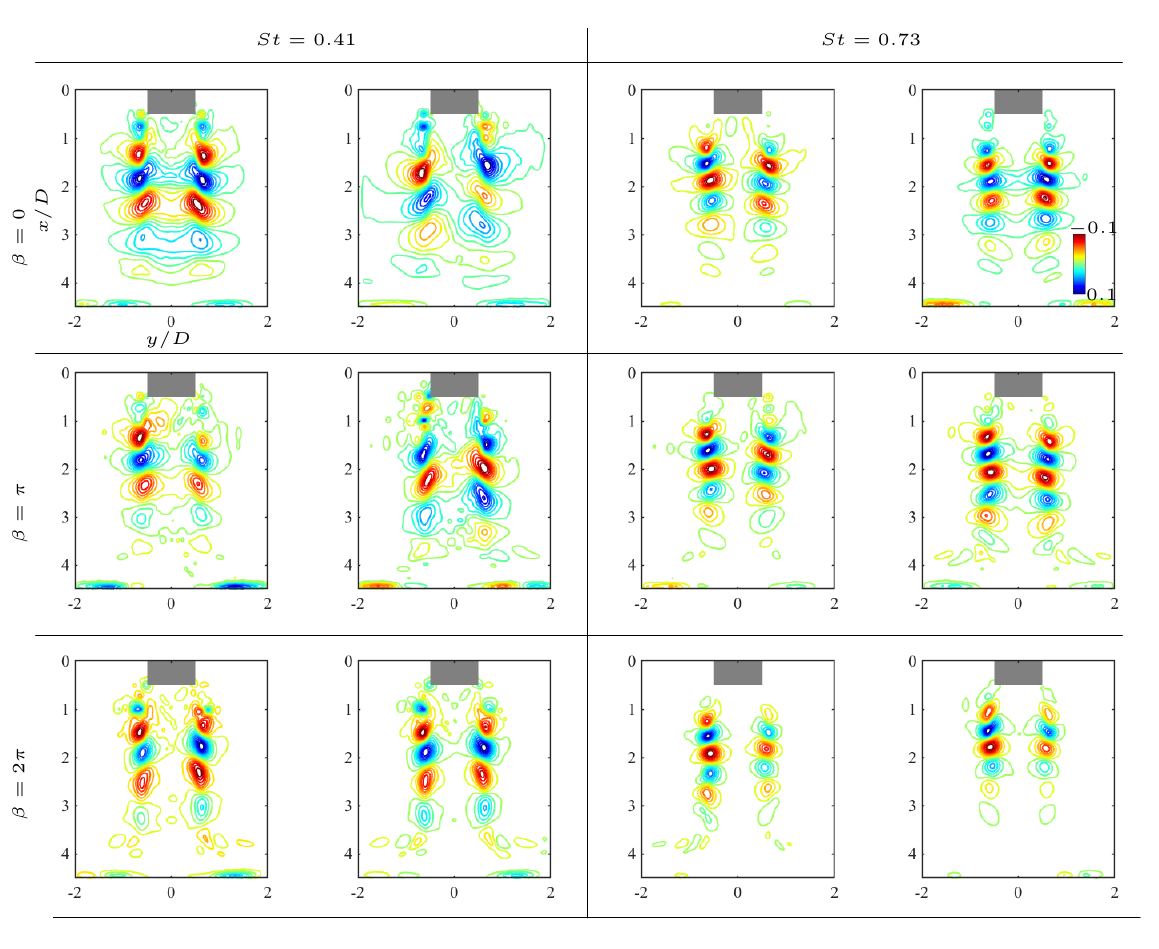}
\caption{Real component of transverse velocity for global modes at $\beta=0$, $\pi$ and $2\pi$.}
\label{fig:globalmodes}
\end{figure}

The time and spanwise averaged impinging jet flow is unstable. 
We perform linear stability analysis of time-averaged flow from $\beta=0$ to $\beta=10\pi$ as described in Section~\ref{sec:Stability}. 
For each spanwise wavenumber $\beta$, we shift the search region sixty positions along with the frequency range of $0 \le St \le 1.25$ and seven values along with the growth rate range of $-0.06 \le \lambda_r/2\pi U_ref \le 0.14$ to cover the eigenvalues of interest. 
Figure~\ref{fig:figure1} shows the eigenspectra at $\beta=0$, $\pi$ and $2\pi$. 
The eigenvalues have positive growth rates, which reveal that the spanwise and time-averaged mean flow is unstable. 
Figure~\ref{fig:globalmodes} shows the modal structure of these unstable modes at $St=0.41$ and $0.73$ as a function of $\beta$ using the real component of the transverse velocity $\hat{v}$.
These modes present distinctively strong structures along the shear layer in the free jet region. 
The structures of the modes gradually vanish at the impinging region as well as the wall jet region. 
This is unsurprising as these modes resemble the Kelvin-Helmholtz instability, which is well known to be the intrinsic instability in the impinging jet flow.
As the frequency increases, the scale of the mode structures changes reversely. 
At a similar frequency range, the global modes exhibit symmetric or antisymmetric structures with different growth rates. 
For example, at $\beta=0$ and frequency $St\sim0.41$, the K-H instability is symmetric for $\lambda_r D/ 2\pi a_\infty=0.1$, while it is anti-symmetric for $\lambda_r D/ 2\pi a_\infty=0.08$.

The existence of both symmetric and anti-symmetric modes however, is not surprising as the previous studies on circular jets~\citep{Powell1953} have shown that the shear layer can support four resonant modes labeled as axisymmetric mode A (divided into A1 and A2 ), flapping mode B, helical mode C, and sinuous mode D.
For the present case, the instantaneous LES flow field shows a strong flapping motion in the jet plume at $H/D=4$, indicating the dominant role of the anti-symmetric modes. 
This dominance however, is likely to vary with the impingement height as seen previously by 
Davis \emph{et al.}~\cite{Davis2015}, where a change in height from $H/D=4$ to $4.5$ resulted in different mode shapes.  
Similar analysis for different heights is being pursued by the authors and will be presented in the final manuscript.

\subsection{Discounted resolvent analysis}
Since the time-averaged flow is temporally unstable, it becomes crucial to highlight amplifications that occur on a shorter time scale than those associated with the asymptotic behavior observed with classical instability theory.
This consideration aids in achieving the main objective of finding preferred energy transfer mechanisms from the mean flow to the fluctuation field, which are necessary to provide physical insight into potential flow control strategies. 
To achieve this objective, the discounting technique ~\cite{jovanovic2004modeling,yeh2018resolvent,liu2021unsteady} is employed to obtain forcing and response modes.
The method introduces a free parameter, denoted the discounting parameter $\kappa$, the choice of which is predicated on information about the most unstable growth rate as obtained from the stability analysis. 
Other methods to address the unstable linear operator may be found in Ref.~\cite{pickering2019eddy}.
Consequently, we substitute the resolvent operator $\mathbb{R}$ by discounted resolvent operator $\mathbb{R'}$ in eqn.\ref{eqn: Resolvent} as follows 
\begin{equation}\label{eqn:disResolvent}
\mathbb{R'}=[\text{i}(\lambda+\text{i}\kappa) \mathbb{I} - \mathbb{L}({\bm{\bar q}},\beta)]^{-1}\mathbb{M}.
\end{equation}
For the present test case, the discounting parameter is chosen as $\kappa=0.159$ based on the global stability analysis results.

\begin{figure}
\centering
\includegraphics[width=0.5\textwidth]{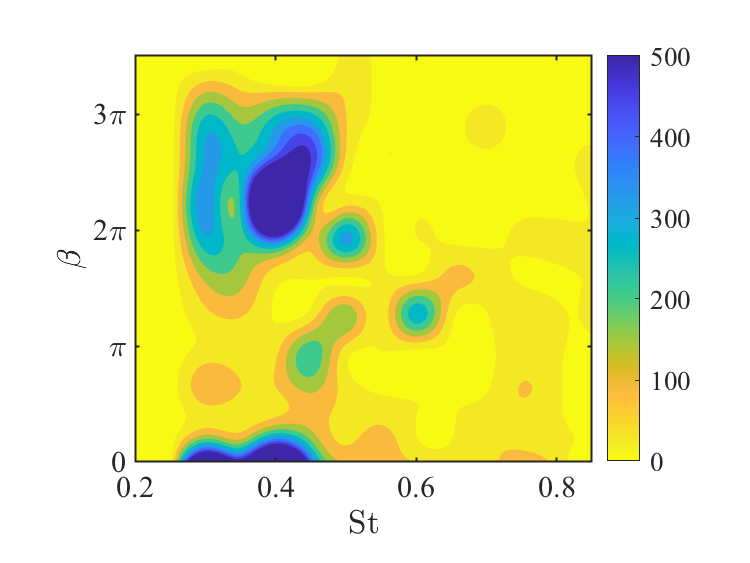}
\caption{The leading energy amplification $\sigma_1$ over $\beta$ and $St$ for a discount parameter of $\kappa=0.14$}
\label{fig:dR1}
\end{figure}
Figure~\ref{fig:dR1} shows the leading singular value as a function of $\beta$ and $St$, revealing the optimal energy amplification to harmonic forcing. 
A large amplification around $0.2<St<0.6$ and $\pi<\beta<3\pi$ is observed from the figure. 
In addition, since the shear layer modes primarily stem from two-dimensional shear-layer instabilities, the gain distribution with $\beta=0$ exhibits discrete peaks at the impingement tones.

\begin{figure}
\centering
\includegraphics[width=0.9\textwidth]{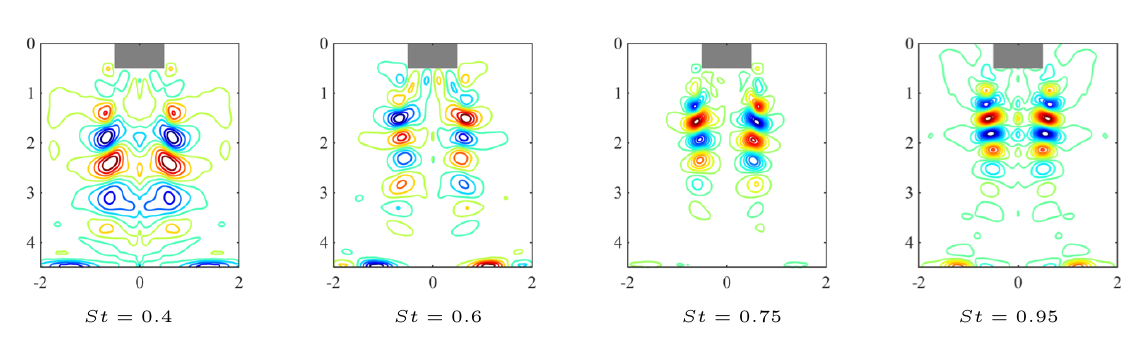}
\caption{Leading response mode at $\beta=\pi$ at frequency $St=0.4$, $0.6$, $0.75$ and $0.95$, illustrating with the real component of $\hat{v}$.}
\label{fig:dR2}
\end{figure}

The representative response modes are presented in Fig.~\ref{fig:dR2} for $\beta=\pi$ at $St=0.4$, 0.6, 0.75 and 0.95 using
the real transverse velocity component. 
Similar to the previous stability analysis, these modes resemble a Kelvin-Helmholtz type of mode structure.  
As expected, with an increase in frequency, the mode structures become smaller and exhibit a smaller wavelength. 
Both symmetric and anti-symmetric response modes are observed. 
As the spanwise wavenumber increases, the response mode structures are confined to the shear-layer region with smaller-scale structures (not shown).

In order to assess the accuracy by which resolvent modes capture the mode shapes, we utilize Spectral Proper Orthogonal Decomposition (SPOD) of the unsteady LES data.
SPOD is a data-driven technique that decomposes the flow-field into a set of orthogonal modes in the frequency domain that optimally capture the flow's energy based on a user-defined norm.
It is well known that SPOD modes are identical to the resolvent modes when the resolvent mode expansion coefficients are uncorrelated, i.e., white-noise forcing~\cite{towne2018spectral}.

SPOD modes aredetermined by solving the following Fredholm integral eigenvalue problem:
\begin{equation}
    \int R \left(\textbf x, \textbf{x'}, \textit{St}\right) W(\textbf{x'}) \phi_n \left( \textbf{x'}\right) \text{d\textbf{x'}} = \lambda_n \left( \textit{St}\right) \phi_n \left(\textbf x,\textit{St}\right).
\end{equation}
Here $\phi_n\left(\textbf x,\textit{St}\right)$ represents the $n^{\text{th}}$ SPOD mode at a particular \emph{St} value and $\lambda_n$ is the energy associated with that mode.
$W$ and $R$ represent the weight and the cross-spectral density matrices respectively.
In order to perform the Fourier transform, the flow-field is arranged in 46 blocks of 256 snapshots with a $75\%$ overlap.
A periodic Hann window is employed to minimize spectral leakage.
Special care is taken to ensure that the conclusions drawn from the SPOD modes are insensitive to these choices.

\begin{figure}
    \centering
    \subfloat[]{\includegraphics[width=0.5\textwidth]{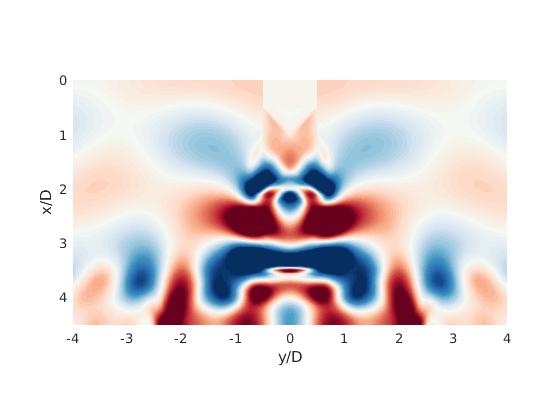}}
    \subfloat[]{\includegraphics[width=0.5\textwidth]{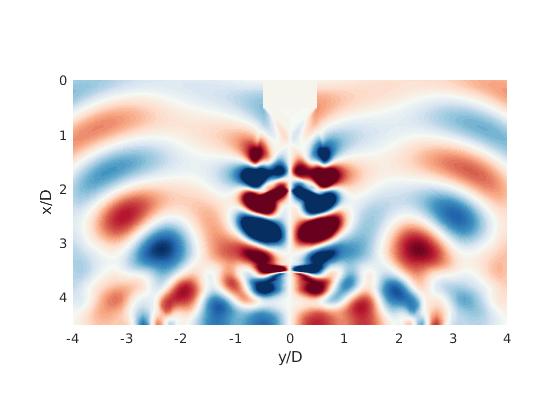}}
    \caption{Leading SPOD pressure modes at (a) $St=0.4$ and (b) $St=0.6$.}
    \label{fig:SPODmodes}
\end{figure}

Figure~\ref{fig:SPODmodes} shows the leading pressure SPOD modes along the jet center plane at $St=0.4$ and $0.6$. 
The mode shapes exhibit a symmetric and anti-symmetric behaviour similar to the resolvent modes at these frequencies as shown previously in Fig.~\ref{fig:dR2} for transverse velocity.
A more comprehensive one-to-one comparison will be presented in the final manuscript.
Nevertheless, this demonstrates the capability of resolvent analysis to accurately represent the flow's response to harmonic forcing. 
This is leveraged next to provide insights into flow control.

\subsection{Flow response to localized component-wise forcing}
A control strategy may be designed based on the information of the response output and forcing input.
The primary forcing modes (not shown) exhibit white noise forcing in the shear layer near the nozzle lip.
As such, forcing around the nozzle lip is a natural choice for flow control, with the potential to significantly modify the behavior of shear layer structures in the free jet region. Inspired by the location control actuation, we consider localization of the input forcing to two locations: (i) the nozzle walls near the nozzle exit $(x/D \leq 0.5)$ and (ii) impinging ground surface $(x/D> 4.1)$.

The component-wise forcing input is selected by prescribing constraints in the coupling matrix $\mathbb{M}$ in eqn.~\ref{eqn: Resolvent}.
The forcing contains five components expressed as $\hat{\bm f}=[\hat{\rho}_f, \hat{u}_f, \hat{v}_f, \hat{w}_f, \hat{T}_f]$ in eqn.~(\ref{eqn:decom_NS}), of which the spanwise velocity component forcing $\hat{w}_f$ is neglected. 
In the coupling matrix $\mathbb{M}$, the elements at the location of interest are set to the chosen component forcing, which localizes the input term $\mathbb{M} \hat{\bm f}$. The matrix $\mathbb{M}$ thus serves two purposes, spatial restriction, and imposition of a component-wise forcing filter. 
The results identify the influence of different types of forcing at these two locations on energy amplification and response structures. 

\begin{figure}
\centering
\includegraphics[width=\textwidth]{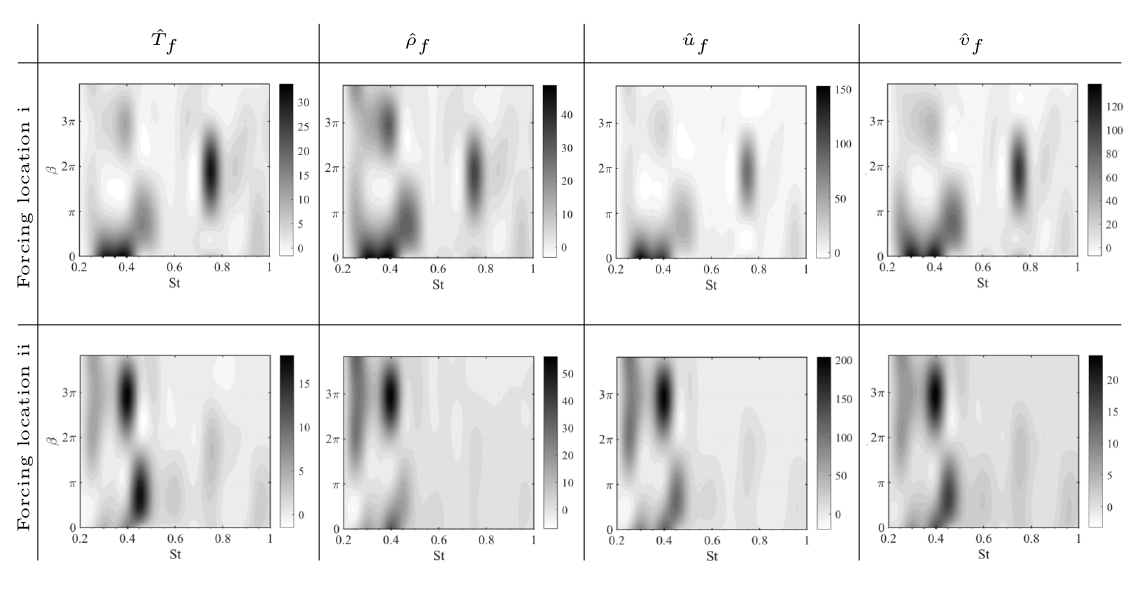}
\caption{Optimal gain of localized-componentwise forcing versus spanwise wavenumbers and frequencies.}
\label{fig:LR1}
\end{figure}

Figure~\ref{fig:LR1} show the energy amplification results with the localized component-wise forcings corresponding to location (i). 
The streamwise velocity forcing results in the strongest energy amplification, followed by the $\hat{v}_f$ , $\hat{\rho}_f$ and $\hat{T}_f$ forcings in that order. 
The localization of $\hat{u}_f$ forcing dramatically reduces the energy amplification to about one-third that with the global forcing shown in Fig.~\ref{fig:dR1}. 
Nonetheless, compared to the global forcing structures, the much smaller localized forcing at a position where an actuator may actually be placed attains substantial energy amplification.
A slightly different trend of energy amplification is observed for the three-dimensional cases ($\beta>0$) with 
another gain peak around $0.6<St<0.8$.

For location (ii), a strong energy amplification is detected in the streamwise velocity forcing followed by the $\hat{\rho}_f$, $\hat{v}_f$, and $\hat{T}_f$ forcings in that order. 
Contrary to the gain maps at location (i), the gain obtained from a three-dimensional forcing around $2\pi<\beta<4\pi$ and $0.2<St<0.6$ is significantly higher than gains obtained at $\beta=0$. 
These gain contours provide insights into the type, wavenumber and the frequency of actuators that can be employed for noise control.
A more complete analysis in underway and will be presented in the final manuscript.

\section{Ongoing work and Conclusion}\label{conclusion}
We perform a large-eddy simulation over a supersonic planar impinging jet at $M=1.27$ and Reynolds number based on nozzle size of 527,000. The frequency tones established between nozzle exit and ground surface agree with the prediction from Powell's formula.

We further adopt global stability and resolvent analysis to identify intrinsic instability. From the global stability analysis, we learn that the time and spanwise averaged flow is unstable. The mode structure indicates the Kelvin-Helmholtz instability as the unique instability existing in the impinging jet flow. Both symmetric and anti-symmetric modes are observed from the global stability analysis. Resolvent analysis is performed over frequency $(St=fU/D)$ and spanwise wavenumber $\beta$ ranges of $0.1\le St \le 1.05$ and $0 \le \beta \le 10\pi$, respectively. The gain-peak resides in the $0.2 \le St \le 0.5$ and $\pi \leq \beta \leq 3\pi$ ranges. We observe that the energy amplification mechanics is dominated by a Kelvin–Helmholtz type of response. At $St\sim 0.4$, this KH response exhibits a symmetric structure at $\beta = 0$, $\pi$, $3\pi$, $4\pi$, $5\pi$ and $6\pi$ with higher energy amplification compared to its anti-symmetric counterparts at other $\beta$-values. The efforts focus on localizing the forcing near the nozzle to consider notional actuators that can separately introduce individual velocity components, density, and temperature forcing, respectively, to target noise mitigation. The result showed a slightly different energy amplification by localized component-wise forcing.

\section*{Acknowledgments}
This work was performed in part under the sponsorship of the Office of Naval Research (Contract N00014-18-1-2506) with Dr. D. Gonzalez serving as Project Monitor. The views and conclusions contained herein are those of the authors and do not represent the opinion of the Office of Naval Research or the U.S. government. Computational resources were provided by the DoD High Performance Computing Modernization Program as well as the Ohio Supercomputer Center. Several figures were made using FieldView software with licenses obtained from the Intelligent Light University Partnership Program. 

\bibliography{bibliography}

\begin{thebibliography}{36}
\newcommand{\enquote}[1]{``#1''}
\providecommand{\natexlab}[1]{#1}
\providecommand{\url}[1]{\texttt{#1}}
\providecommand{\urlprefix}{URL }
\expandafter\ifx\csname urlstyle\endcsname\relax
  \providecommand{\doi}[1]{\discretionary{}{}{}https://doi.org/#1}\else
  \providecommand{\doi}[1]{\discretionary{}{}{}\urlstyle{rm}\url{https://doi.org/#1}}\fi

\bibitem[{Powell(1988)}]{powell1988sound}
Powell, A., \enquote{The sound-producing oscillations of round underexpanded
  jets impinging on normal plates,} \emph{The Journal of the Acoustical Society
  of America}, Vol.~83, No.~2, 1988, pp. 515--533.

\bibitem[{Edgington-Mitchell(2019)}]{edgington2019aeroacoustic}
Edgington-Mitchell, D., \enquote{Aeroacoustic resonance and self-excitation in
  screeching and impinging supersonic jets--a review,} \emph{International
  Journal of Aeroacoustics}, Vol.~18, No. 2-3, 2019, pp. 118--188.

\bibitem[{Prasad et~al.(2021{\natexlab{a}})Prasad, Stahl, and
  Gaitonde}]{prasad2021exchange}
Prasad, C., Stahl, S., and Gaitonde, D., \enquote{Exchange Mechanisms between
  Hydrodynamic and Acoustic Components of an Under-expanded Supersonic
  Impinging Jet,} \emph{AIAA AVIATION 2021 FORUM}, 2021{\natexlab{a}}, p. 2118.

\bibitem[{Krothapalli et~al.(1999)Krothapalli, Rajkuperan, Alvi, and
  Lourenco}]{krothapalli1999flow}
Krothapalli, A., Rajkuperan, E., Alvi, F., and Lourenco, L., \enquote{Flow
  field and noise characteristics of a supersonic impinging jet,} \emph{4th
  AIAA/CEAS aeroacoustics conference}, 1999, p. 2239.

\bibitem[{Worden et~al.(2013-2187)Worden, Gustavsson, Shih, and
  Alvi}]{worden2013acoustic}
Worden, T., Gustavsson, J., Shih, C., and Alvi, F.~S., \enquote{Acoustic
  measurements of high-temperature supersonic impinging jets in multiple
  configurations,} \emph{19th AIAA/CEAS Aeroacoustics Conference}, 2013-2187.

\bibitem[{Brehm et~al.(2016)Brehm, Housman, and Kiris}]{brehm2016noise}
Brehm, C., Housman, J.~A., and Kiris, C.~C., \enquote{Noise generation
  mechanisms for a supersonic jet impinging on an inclined plate,}
  \emph{Journal of Fluid Mechanics}, Vol. 797, 2016, pp. 802--850.

\bibitem[{Tsutsumi et~al.(2015)Tsutsumi, Ishii, Ui, Tokudome, and
  Wada}]{tsutsumi2015study}
Tsutsumi, S., Ishii, T., Ui, K., Tokudome, S., and Wada, K., \enquote{Study on
  acoustic prediction and reduction of {Epsilon} launch vehicle at liftoff,}
  \emph{Journal of Spacecraft and Rockets}, 2015.

\bibitem[{Prasad et~al.(2021{\natexlab{b}})Prasad, Yenigelen, and
  Morris}]{prasad2021effectlaunchpad}
Prasad, C., Yenigelen, E., and Morris, P.~J., \enquote{Effect of Launchpad
  Modification on the Hydrodynamic and Acoustic Modes of an Impinging Jet,}
  \emph{AIAA Scitech 2021 Forum}, 2021{\natexlab{b}}, p. 1417.

\bibitem[{Alvi et~al.(2003)Alvi, Shih, Elavarasan, Garg, and
  Krothapalli}]{alvi2003control}
Alvi, F.~S., Shih, C., Elavarasan, R., Garg, G., and Krothapalli, A.,
  \enquote{Control of supersonic impinging jet flows using supersonic
  microjets,} \emph{AIAA journal}, Vol.~41, No.~7, 2003, pp. 1347--1355.

\bibitem[{Norum(2004)}]{norum2004reductions}
Norum, T., \enquote{Reductions in multi-component jet noise by water
  injection,} \emph{10th AIAA/CEAS Aeroacoustics Conference}, 2004, p. 2976.

\bibitem[{Fukuda et~al.(2011-2814)Fukuda, Tsutsumi, Shimizu, Takaki, and
  Ui}]{fukuda2011examination}
Fukuda, K., Tsutsumi, S., Shimizu, T., Takaki, R., and Ui, K.,
  \enquote{Examination of sound suppression by water injection at lift-off of
  launch vehicles,} \emph{17th AIAA/CEAS Aeroacoustics Conference (32nd AIAA
  Aeroacoustics Conference)}, 2011-2814.

\bibitem[{Salehian et~al.(2018-0519)Salehian, Kourbatski, Golubev, and
  Mankbadi}]{salehian2018numerical}
Salehian, S., Kourbatski, K., Golubev, V.~V., and Mankbadi, R.~R.,
  \enquote{Numerical Aspects of Rocket Lift-off Noise with Launch-Pad Aqueous
  Injection,} \emph{2018 AIAA Aerospace Sciences Meeting}, 2018-0519.

\bibitem[{Liu et~al.(2018)Liu, Sun, Cattafesta, Ukeiley, and
  Taira}]{liu2018resolvent}
Liu, Q., Sun, Y., Cattafesta, L.~N., Ukeiley, L.~S., and Taira, K.,
  \enquote{Resolvent Analysis of Compressible Flow over a Long Rectangular
  Cavity,} {AIAA} Paper 2018-0588, 2018.

\bibitem[{Liu et~al.(2021)Liu, Sun, Yeh, Ukeiley, Cattafesta, and
  Taira}]{liu2021unsteady}
Liu, Q., Sun, Y., Yeh, C.-A., Ukeiley, L.~S., Cattafesta, L.~N., and Taira, K.,
  \enquote{Unsteady control of supersonic turbulent cavity flow based on
  resolvent analysis,} \emph{Journal of Fluid Mechanics}, Vol. 925, 2021.

\bibitem[{Bhargav et~al.(2021)Bhargav, Song, Sellappan, Alvi, and
  Kumar}]{Bhargav2021}
Bhargav, V.~N., Song, M., Sellappan, P., Alvi, F.~S., and Kumar, R.,
  \enquote{{Experimental Characterization of Supersonic Single- and
  Dual-Impinging Jets},} \emph{AIAA Journal}, 2021, pp. 1--17.
\newblock \doi{10.2514/1.j059687}.

\bibitem[{Davis et~al.(2015)Davis, Edstrand, Alvi, Cattafesta, Yorita, and
  Asai}]{Davis2015}
Davis, T., Edstrand, A., Alvi, F., Cattafesta, L., Yorita, D., and Asai, K.,
  \enquote{{Investigation of impinging jet resonant modes using unsteady
  pressure-sensitive paint measurements},} \emph{Experiments in Fluids},
  Vol.~56, No.~5, 2015, pp. 1--13.
\newblock \doi{10.1007/s00348-015-1976-9}.

\bibitem[{Kumar et~al.(2013)Kumar, Wiley, Venkatakrishnan, and
  Alvi}]{Kumar2013}
Kumar, R., Wiley, A., Venkatakrishnan, L., and Alvi, F., \enquote{{Role of
  coherent structures in supersonic impinging jets},} \emph{Physics of Fluids},
  Vol.~25, No.~7, 2013.
\newblock \doi{10.1063/1.4811401}.

\bibitem[{Roe(1981)}]{roe1981approximate}
Roe, P.~L., \enquote{Approximate Riemann solvers, parameter vectors, and
  difference schemes,} \emph{Journal of computational physics}, Vol.~43, No.~2,
  1981, pp. 357--372.

\bibitem[{Van~Leer(1979)}]{van1979towards}
Van~Leer, B., \enquote{Towards the ultimate conservative difference scheme. V.
  A second-order sequel to Godunov's method,} \emph{Journal of computational
  Physics}, Vol.~32, No.~1, 1979, pp. 101--136.

\bibitem[{Pulliam and Chaussee(1981)}]{pulliam1981diagonal}
Pulliam, T.~H., and Chaussee, D., \enquote{A diagonal form of an implicit
  approximate-factorization algorithm,} \emph{Journal of Computational
  Physics}, Vol.~39, No.~2, 1981, pp. 347--363.

\bibitem[{Beam and Warming(1978)}]{beam1978implicit}
Beam, R.~M., and Warming, R., \enquote{An implicit factored scheme for the
  compressible Navier-Stokes equations,} \emph{AIAA journal}, Vol.~16, No.~4,
  1978, pp. 393--402.

\bibitem[{Bogey and Bailly(2010)}]{bogey_bailly_2010}
Bogey, C., and Bailly, C., \enquote{Influence of nozzle-exit boundary-layer
  conditions on the flow and acoustic fields of initially laminar jets,}
  \emph{Journal of Fluid Mechanics}, Vol. 663, 2010, p. 507–538.

\bibitem[{Nataraj~Bhargav et~al.(2020)Nataraj~Bhargav, Song, Sellappan, Alvi,
  and Kumar}]{nataraj2020unsteady}
Nataraj~Bhargav, V., Song, M., Sellappan, P., Alvi, F.~S., and Kumar, R.,
  \enquote{Unsteady Characteristics of Resonant Supersonic Dual Impinging
  Jets,} \emph{AIAA Scitech 2020 Forum}, 2020.

\bibitem[{Stahl et~al.(2021)Stahl, Prasad, and
  Gaitonde}]{stahl2021distinctions}
Stahl, S.~L., Prasad, C., and Gaitonde, D.~V., \enquote{Distinctions between
  single and twin impinging jet dynamics,} \emph{The Journal of the Acoustical
  Society of America}, Vol. 150, No.~2, 2021, pp. 734--744.

\bibitem[{Powell(1953)}]{Powell1953}
Powell, A., \enquote{On edge tones and associated phenomena,} \emph{Acustica},
  Vol.~3, 1953, pp. 233--43.

\bibitem[{Farrell and Ioannou(1993)}]{farrell1993stochastic}
Farrell, B.~F., and Ioannou, P.~J., \enquote{Stochastic forcing of the
  linearized Navier--Stokes equations,} \emph{Physics of Fluids A: Fluid
  Dynamics}, Vol.~5, No.~11, 1993, pp. 2600--2609.

\bibitem[{McKeon and Sharma(2010)}]{mckeon2010critical}
McKeon, B.~J., and Sharma, A.~S., \enquote{A critical-layer framework for
  turbulent pipe flow,} \emph{J. Fluid Mech.}, Vol. 658, 2010, pp. 336--382.

\bibitem[{Liu et~al.(2016)Liu, G{\'o}mez, and Theofilis}]{liu2016linear}
Liu, Q., G{\'o}mez, F., and Theofilis, V., \enquote{Linear instability analysis
  of low-${R}e$ incompressible flow over a long rectangular finite-span open
  cavity,} \emph{Journal of Fluid Mechanics}, Vol. 799, 2016.

\bibitem[{Liu and Gaitonde(2021)}]{liu2021acoustic}
Liu, Q., and Gaitonde, D., \enquote{Acoustic response of turbulent cavity flow
  using resolvent analysis,} \emph{Physics of Fluids}, Vol.~33, No.~5, 2021, p.
  056102.

\bibitem[{Jovanovi\'{c} and Bamieh(2005)}]{jovanovic2005componentwise}
Jovanovi\'{c}, M.~R., and Bamieh, B., \enquote{Componentwise energy
  amplification in channel flows,} \emph{J. Fluid Mech.}, Vol. 534, 2005, pp.
  145--183.

\bibitem[{Schmid and Henningson(2012)}]{schmid2012stability}
Schmid, P.~J., and Henningson, D.~S., \emph{Stability and transition in shear
  flows}, Springer, 2012.

\bibitem[{Chu(1965)}]{chu1965energy}
Chu, B.-T., \enquote{On the energy transfer to small disturbances in fluid flow
  (Part {I}),} \emph{Acta Mechanica}, Vol.~1, No.~3, 1965, pp. 215--234.

\bibitem[{Jovanovi\'{c}(2004)}]{jovanovic2004modeling}
Jovanovi\'{c}, M.~R., \enquote{Modeling, analysis, and control of spatially
  distributed systems,} Ph.D. thesis, University of California, Santa Barbara,
  2004.

\bibitem[{Yeh and Taira(2019)}]{yeh2018resolvent}
Yeh, C.-A., and Taira, K., \enquote{Resolvent-analysis-based design of airfoil
  separation control,} \emph{J. Fluid Mech.}, Vol. 867, 2019, pp. 572--610.

\bibitem[{Pickering et~al.(2019-2454)Pickering, Rigas, Sipp, Schmidt, and
  Colonius}]{pickering2019eddy}
Pickering, E.~M., Rigas, G., Sipp, D., Schmidt, O.~T., and Colonius, T.,
  \enquote{Eddy viscosity for resolvent-based jet noise models,} \emph{AIAA
  Paper No.}, 2019-2454.

\bibitem[{Towne et~al.(2018)Towne, Schmidt, and Colonius}]{towne2018spectral}
Towne, A., Schmidt, O.~T., and Colonius, T., \enquote{Spectral proper
  orthogonal decomposition and its relationship to dynamic mode decomposition
  and resolvent analysis,} \emph{Journal of Fluid Mechanics}, Vol. 847, 2018,
  pp. 821--867.

\end{thebibliography}

\end{document}